\documentstyle{article}

\def\b{\bibitem}
\def\di{(\frac{\sigma}{\theta})_{0}}
\def\ro{(\frac{\omega}{\theta})_{0}}
\def\p{\cdot}
\def\xu{\frac{1}{x}}
\def\xd{\frac{2}{x}}
\def\curv{\stackrel{3}{R}}
\def\bpiu{\beta_{+}}

\def\K{\Gamma}

\def\v0{\Delta}
\def\C{\Sigma}
\def\mp{M_{PL}}
\def\tt{(1-\tau^{2})}
\def\bl{\beta_{LIM}}
\def\tetacr{\theta_{\star}}
\def\bs{\beta_{\star}}

\title{Inflation for Bianchi IX models}
\author{
Roberto Bergamini\thanks{
bergamini@astbo1.bo.cnr.it}
 \and 
Paolo Sedici 
\and 
Paolo Verrocchio\thanks{
verrocchio@astbo1.bo.cnr.it}
}
\date{}

\begin{document}

\maketitle
\begin{center}
{\small\it
Istituto di Radioastronomia del C.N.R.,via Gobetti 101, 40100,
 Bologna, Italy}
\end{center}

\bigskip

\begin{abstract}

\noindent
The influence of Inflation on initial (i.e. at Planck's epoch) large
anisotropy of the Universe is studied. 
To this end we consider a more general metric than the isotropic one: 
the locally rotationally symmetric (L.R.S.) Bianchi IX metric. 
We find, then, a large set of initial
conditions of intrinsic curvature and shear allowing 
an inflationary epoch that make the anisotropy negligible.
These are not trivial because of the non-linearity of the 
Einstein's equations. 
\end{abstract}

\newpage
\normalsize
\centerline{\bf 1. Introduction}

\noindent
The observations of Cosmic Microwave Background Radiation can be
utilized to give some constraints on the anisotropy of the Universe.

\noindent
These limits have been derived analytically by
Hawkings and Collins \cite{hawk_coll}, considering little deviations 
from isotropy; whilst   
Barrow, Juszkiewicz and Sonoda \cite{ba_ju_so} have computed the 
temperature pattern and the angular correlation function for the temperature 
perturbations expected in anisotropic models. 
Bunn, Ferreira and Silk have used the theoretical temperature pattern of the 
Bianchi model of type ${VII_{h}}$ \cite{bu_ferr_silk} to determinate 
the values of shear and vorticity making a best-fit with 
C.M.B.R. experimental data.  
A model independent approach at the problem has been introduced by 
Maartens, Ellis and Stoeger \cite{maa_ell_sto}.

\noindent
All this results are summarized in table (\ref{limiti}) and they show that,
actually, Universe is isotropic with a good approximation.

\noindent
The set of initial conditions allowed from General Relativity is 
very larger than isotropy; so we must look for a physical process
making Universe isotropic if, at Planck's epoch, it was highly anisotropic.

\noindent
The theory of Inflation, as it 
leads to `natural' prediction about the value of the curvature of the Universe,
about the spectrum of scalar and tensor perturbation 
\cite{li_li,liddle}, and it solves the topological defects, flatness 
and horizon problems \cite{hu_tu_we}, 
could be this physical process; even if it is not the only candidate
\cite{barr1,misner,misner2,misner3,haw_lut,ber_gia,be_ka_li}.

\noindent
To verify this hypothesis we assume an anisotropic metric, the Bianchi metric
\cite{maccallum,ell_mac}, and we introduce the stress-energy tensor
of a scalar field minimally coupled to gravity that can give an
Inflationary epoch \cite{guth,linde}.

\noindent
In this way,
we are looking for the initial conditions allowing Inflation and study 
the evolution of anisotropy, comparing final values with the observed one.

\noindent
Between the different Bianchi types, we studied Bianchi IX model,
because it is the only one allowing positive
intrinsic curvature ${\curv}$ \cite{stephani}.

\noindent
Using the Raychaudhuri relation \footnote{ We will use such units 
that $c = h = k = 1, \frac{8\pi G}{3c^{2}} = \mp^{-2}$.}

\begin{equation}
\frac{2}{3}\theta^{2} = 2\frac{V(\phi)}{\mp^{2}} +2\sigma^{2}-\curv
\label{vincolo}
\end{equation}

\noindent
(where $V(\phi)$ is the potential energy of scalar field,
$\theta$ measures the rate of expansion of the Universe and $\sigma$
is the shear of the homogeneous hypersurfaces \cite{ellis})

\noindent
it is clear that,
in Bianchi IX model, the positive value of $\curv$, 
describing a closed Universe, could cancel the expansion of the volume of the
Universe, even with a scalar field acting as cosmological 
constant \cite{wald,ro_el}; whereas in the other Bianchi types, open 
or flat, shear and intrinsic curvature `help' the expansion.

\noindent
The study of evolution of shear   
in Bianchi V model \cite{he_me,ma_jo}\ shows that 
Inflation leads this model to be completely isotropic.

\noindent
An attempt to study the dynamics of Bianchi models with a scalar field,
using the N$\ddot{\rm o}$ther symmetries in minisuperspace, 
has been recently performed \cite{capozziello}; 
but there is not yet a complete analysis.

\vspace{.5cm}
\centerline{\bf 2. Inflation with Bianchi IX metric}

\noindent 
Let us consider the metric of locally rotationally symmetric (L.R.S.) 
Bianchi IX Universe \footnote{
The latin index i,j,etc =1,2,3.}: 
\cite{misner4,taub,ry_sh}:

\begin{equation}
ds^{2} = -\mp^{-2}N^{2}(\lambda)d\lambda^{2}+e^{-2\alpha}[e^{-2\beta}
({{\bf \omega}^{1}})^{2}+e^{\beta}(({{\bf \omega}^{2}})^{2}+
({{\bf \omega}^{3}})^{2})]
\label{metrica}
\end{equation}

\noindent
where $N$ is the lapse function and the one-forms ${\bf \omega}$ are defined
by: 

\begin{equation}
d{\bf \omega}^{i} = \frac{1}{2}\epsilon^{i}_{jk}{\bf \omega}^{j}\wedge
{\bf \omega}^{k}
\label{struttura}
\end{equation}

\noindent
The evolution's rate of $\beta$ is related to shear:

\begin{equation}
\label{shear} 
\sigma^{2} = 3 \left(
\frac{\mp}{N} 
\beta^{\prime}
\right)^{2}
\end{equation}

\noindent
The general Bianchi IX model is more complicated with respect to L.R.S.
Bianchi IX model; in fact in that case the Einstein's equations
have not been yet solved. 
Our simplification does not affect the 
fundamental feature of the model that we want to study:
the effect of positive curvature, and it allows us to
compute the anisotropic initial conditions compatible with Inflation.

\noindent 
The Lagrangian for gravitational and scalar field is 
\cite{adm,ryan,ug_ja_ro}:

\begin{equation}
\label{lagtotale}
L = \mp^{2} \left(
\frac{1}{2}w(-{\alpha^{\prime}}^{2}+{\beta^{\prime}}^{2}+
\frac{1}{\mp^{2}}{\phi^{\prime}}^{2})-
\frac{12}{w}e^{-4\alpha}U(\beta) - \frac{24}{w}e^{-6\alpha}
\frac{V(\phi)}{\mp^4} \right)     
\end{equation}

\begin{equation}
\label{potgrav}
U(\beta) = -2e^{-2\beta}+\frac{1}{2}e^{-8\beta}  \\
\end{equation}

\noindent
where the gauge freedom is hidden in the function ${w\equiv\frac{12e^{-3\alpha}}
{N(\lambda)}}$.

\noindent
The common feature to different theories of Inflation is the so-called
`slow-roll' approximation \cite{linde3}
in the description of motion of the field.
This implies that Klein-Gordon equation reduces to:

\begin{equation}
3\frac{\mp^2}{N^2}\alpha^{\prime}\phi^{\prime} = 
\frac{\partial V}{\partial \phi}
\label{klein}
\end{equation}

\noindent
With this approximation, models of Inflation satisfy two other conditions on
potential $V(\phi)$:

\begin{equation}
\label{epsilon}
\epsilon \equiv \frac{\mp^{2}}{2} \left(
\frac{1}{V}\frac{\partial V}{\partial \phi} \right)^{2} \ll 1 
\end{equation}

\begin{equation}
\label{eta}
\eta \equiv \mp^{2} \frac{1}{V}
\frac{\partial^{2}V}{\partial \phi^{2}} \ll 1
\end{equation}

\noindent
As we are not interested in the length of Inflation but to initial
conditions causing Inflation, it is enough to verify that the characteristic
time in the evolution of scalar field $T_{\phi}$ is much greater 
than gravitational characteristic time $T_{G}$.

\noindent 
In our model $T_{G} \propto \mp^{-1}$, whereas from assumptions 
(\ref{klein}) and (\ref{epsilon}) 
we obtain $T_{\phi} \propto \left( \frac{\mp^{3}}{\epsilon V(\phi)} \right)$

\noindent
We shall assume a value of about $\mp^{4}$ for energy potential 
density at Planck's epoch in `chaotic' Inflation theory. 
This assumption is justified by quantum effects giving 
radiative corrections to effective potential \cite{na_pa} of this typical 
order of magnitude \cite{lectures}, then:

\begin{equation}
\label{tempiscala}
T_{\phi} = \frac{1}{\epsilon} T_{G}
\end{equation}

\noindent 
We can neglect, then, the `kinetic' contribution of scalar field and assume
that the potential term acts as a cosmological constant.

\noindent
Hence, the contribution to lagrangian due to scalar field becomes:

\begin{equation}
L_{\phi} \approx \mp^{2}\frac{24}{w}e^{-6\alpha}
\end{equation}

\noindent
The solutions of the Einstein's equations in this case have been found by
Cahen and Defrise \cite{ca_de}; following the work of Uggla,Jantzen and
Rosquist we made a different choice of slicing gauge and we have found
the solutions of Einstein's equations for the new functions:

\begin{equation}
W=e^{-3\alpha-3\beta}, Z=e^{2\alpha-2\beta}
\label{trasformazione}
\end{equation}

\noindent
They can be written in the following way:

\begin{equation}
\label{zeta}
Z(\tau) = \K\tt  
\end{equation}

\begin{equation}
\label{vi}
W(\tau) = \K^{-\frac{5}{2}}
\left\{
\left(\K-\frac{4}{3} \right)
\left[8\tt^{\frac{1}{2}}-4\tt^{-\frac{1}{2}} \right]  
+\frac{4}{3}\tt^{-\frac{3}{2}} +\C\tau
\right\}
+\v0
\end{equation}

\noindent
where $\K$, $\C$ and $\v0$ are constants of integration.

\noindent
Inserting the solutions (\ref{zeta})(\ref{vi}) in (\ref{vincolo}),
we obtain the constraint $\v0=0$.

\noindent
The new coordinate time $\tau$ is related to comoving time $t$ by: 

\begin{equation}
dt=-2\mp^{-1}\K^{\frac{1}{2}}W^{-\frac{1}{2}}Z^{-\frac{7}{4}}d\tau
\label{tempoproprio}
\end{equation}

\noindent
We see that when $\tau \rightarrow 1$ then $t\rightarrow -\infty$, and
when $\tau \rightarrow -1$ then $t\rightarrow \infty$.

\noindent
We choose as initial comoving time  
$t_{PL}=10^{-43}$ sec.; the correspondence with initial coordinate 
time $\tau_{0}$, for a given
solution of Einstein equations, depends on the choice of integration
constant in (\ref{tempoproprio}).

\noindent
The physically important part of the solutions is limited to interval 
[$\tau_{0}$,$-1$[, i.e. [$t_{PL}$, $+\infty$[.

\noindent
The physical quantities $\theta$ and $\sigma$ can be expressed
as functions of $W$ and $Z$:

\begin{equation}
\label{theta}
\theta = \mp\K^{-\frac{1}{2}}\left(\frac{3}{8}W^{\frac{1}{2}}
Z^{\frac{3}{4}}\frac{dZ}{d\tau} - \frac{1}{4}W^{-\frac{1}{2}}Z^{\frac{7}{4}}
\frac{dW}{d\tau}\right)       
\end{equation}

\begin{equation}
\label{sigma}
\sigma =  \mp\K^{-\frac{1}{2}}\left|\frac{\sqrt{3}}{8}W^{\frac{1}{2}}
Z^{\frac{3}{4}}\frac{dZ}{d\tau} + \frac{\sqrt{3}}{12}W^{-\frac{1}{2}}
Z^{\frac{7}{4}}\frac{dW}{d\tau}\right|   
\end{equation}

\vspace{.5cm}
\centerline{\bf 3. Initial conditions of the Universe}

\noindent
The initial coordinate time chosen is $\tau_{0}=0$; in this way 
by (\ref{trasformazione}) and (\ref{theta}) we can write explicitly
the relation between 
integration constants and the quantities ${\beta}_{0}$ and $\theta_{0}$:

\begin{equation}
\K = \frac{4}{4-e^{-6\beta_{0}}}
\label{condi1}
\end{equation}

\begin{equation}
\C = -\frac{8\theta_{0}}{\mp} \sqrt{\K-1}  
\label{condi2}
\end{equation}

\noindent
Because of (\ref{zeta}) and (\ref{trasformazione}),
$\K$ must be positive.

\noindent 
Equation (\ref{condi1}) implies:

\begin{equation}
\beta_{0} > -\frac{1}{6}\ln (4) \equiv \bl 
\label{blimite}
\end{equation}

\noindent
that, for the equality 

\begin{equation}
\label{intrinseca}
\curv = -\mp^{2}e^{2\alpha}U(\beta)
\end{equation}

\noindent 
is equivalent to:

\begin{equation}
\curv_{0} > 0
\label{positiva}
\end{equation}

\noindent
The volume of the Universe is given by:

\[
{\rm Vol} = e^{-3\alpha}=\frac{W^{\frac{1}{2}}}{Z^{\frac{3}{4}}}
\]

\begin{equation}
= \frac{\sqrt{
\left( \Gamma-\frac{4}{3} \right)
\left[8\tt^{\frac{1}{2}}-4\tt^{-\frac{1}{2}} \right]
+\frac{4}{3}\tt^{-\frac{3}{2}}+\Sigma \tau}}{\Gamma^{2}\tt^{\frac{3}{4}}}  
\label{volume}
\end{equation}

\noindent
The behaviour of this function depends on the two
independent initial conditions $\beta_{0}$ and $\theta_{0}$. The other two
initial conditions $\alpha_{0}$ and $\sigma_{0}$ are related to 
$\beta_{0}$ and $\theta_{0}$ by:

\begin{equation}
\label{altra}
-\alpha_{0} = -\beta_{0}+\frac{1}{2}\ln{\K} 
\end{equation}

\begin{equation}
\label{equipartizione}
\sigma_{0} = \frac{\sqrt{3}}{3} \theta_{0}  
\end{equation}

\noindent
The former is not physically relevant because
simply fixes the length scales at $\tau=0$. 
The latter, instead, is the consequence 
of Hamiltonian constraint (\ref{vincolo}), and it means 
that (with the choice $\tau_{0}=0$)
our model describes an initial `equipartition'
of energy among the different 
gravitational degrees of freedom $\alpha$ e $\beta$. This is
a more general and 'natural' initial condition than isotropy, and it
has been recently analyzed by Barrow \cite{barr2}.

\noindent
The function (\ref{volume}), for different choices of free
parameters $\K$ e $\C$, gives
two possible evolution of the Universe's volume. 
In the first case, the anisotropy cannot stop expansion, and we have;

\begin{equation}
\label{inflasi}
{\rm Vol} = e^{-3\alpha} \rightarrow +\infty ~~{\rm when}~~ \tau \rightarrow -1
\end{equation}

\noindent
In the second case, the volume reaches a maximum and then it shrinks again,
while shear and intrinsic curvature diverge; this happens at a 
coordinate time $\tau_{C} \in [0,-1[$, corresponding always to a finite
comoving time $t_{C}$. Integrating numerically the equation 
(\ref{tempoproprio}) we can show that $t_{C} \approx$ a few $t_{PL}$, and the
exact value depends on initial condition $\K$ and $\C$.

\noindent
The former possibility occurs if $\theta_{0}>\tetacr$, where:

\[
\tetacr \approx 54.60 \mp ~~~~~~~~~~~\beta_{0} \in ]\bl,.996\bl] 
\]

\begin{equation}
\tetacr \approx 0.05\mp \left( \ln
\frac{\bl}{\bl-\beta_{0}} \right)^{3.80} \beta_{0} 
\in [.996\bl,.792\bl]
\label{iniziali}
\end{equation}

\[
\tetacr \approx 0 ~~~~~~~~~~~~~~~~\beta_{0} \in [.792\bl,+\infty[
\]

\noindent
otherwise Universe does not inflate.

\noindent
Hence, it exits an initial value of $\beta_{0} = 0.792\bl \equiv \bs$ 
distinguishing two different behaviours:

\vspace{.5cm}
\noindent
a) $\beta_{0} > \bs$

\noindent
There is Inflation with any initial value of expansion $\theta_{0}$ 
and shear $\sigma_{0}$ , 
even if their values are much greater than $\mp$.  

\noindent
To explain this unexpected result, let us come back 
to Lagrangian (\ref{lagtotale}), obtaining the differential equation for
the evolution of $\beta$ with respect to comoving
time $t$:

\begin{equation}
\frac{d^{2}\beta}{dt^{2}}-3\frac{d\alpha}{dt}\frac{d\beta}{dt}
+\frac{1}{12}\mp^{2}e^{2\alpha}\frac{\partial}{\partial \beta}U(\beta)=0
\label{ripidita}
\end{equation}

\noindent
then, we can show that, with high values of $\theta_{0}$ and $\sigma_{0}$, 
the `friction' term:

\begin{equation}
\label{attrito}
-3\frac{d\alpha}{dt}\frac{d\beta}{dt}
\end{equation}

\noindent
due to combined action of expansion and shear is more relevant 
than `forcing' term:

\begin{equation}
\label{forza}
+\frac{1}{12}\mp^{2}e^{2\alpha}\frac{\partial}{\partial \beta}U(\beta)
\end{equation}

\noindent
because the potential $U(\beta)$ is not enough steep, i.e. 
the value of derivative of potential with respect 
to $\beta$ is not enough large; moreover the large value of $\theta$ implies
that $e^{2\alpha}$ becomes `small' very quickly.

\noindent
After a few Planck's times
the intrinsic curvature and the shear become dynamically negligible 
and they decay:

\begin{equation}
\label{decadimento1}
\sigma^{2} = {\rm Vol}^{-2} 
\end{equation}

\begin{equation}
\label{decadimento2}
\curv = {\rm Vol}^{-\frac{2}{3}} 
\end{equation}

\noindent
whereas expansion $\theta$ reaches his isotropic value $\sqrt{3}\mp$ 
(equation(\ref{theta})).

\vspace{.5cm}
\noindent
b) $\beta_{0} < \bs$

\noindent
This case is more complicated; in fact $\tetacr \neq 0$, 
and it reaches the value of a few ten of Planck's mass when $\beta_{0}
\rightarrow \bl$.

\noindent
Universe inflate if $\theta_{0} > \tetacr$, or, because of 
(\ref{equipartizione}), if $\sigma_{0} > \frac{\sqrt{3}}{3}\tetacr$.
Hence, with respect to previous case, Inflation can be avoided; but,
surprisingly, only with the smallest value of shear.

\noindent
If there is Inflation, equations (\ref{decadimento1})
(\ref{decadimento2}) remain valid; again Universe becomes isotropic.
If, instead, there is not Inflation, we have that:

\begin{equation}
\label{beta2}
\bpiu \rightarrow \infty
\end{equation}

\noindent
The equations (\ref{potgrav})(\ref{sigma})(\ref{intrinseca})
(\ref{beta2}), then, show that:

\begin{equation}
\label{divergenza1}
\sigma^{2} \rightarrow \infty 
\end{equation}

\begin{equation}
\label{divergenza2}
\curv \rightarrow \infty
\end{equation}

\noindent
This behaviour is due to initial large `steepness' of potential $U(\beta)$ when
$\beta_{0} < \bs$. This gives to $\beta$ an initial 'acceleration` enough
to win the 'friction' term (\ref{attrito}) and reach the 
values $\beta \gg 1$

\noindent
In our model, the dependence of evolution of the volume on 
the initial value of intrinsic curvature $\curv_{0}$ is non trivial. 
For example, the relation (\ref{intrinseca}) shows that when 
$\beta_{0} \rightarrow \bl$ the curvature $\curv_{0} \rightarrow 0$, whereas if
$\beta_{0} \rightarrow \infty$ then $\curv_{0} \rightarrow 2\mp^{2}$. 
In the first case, where Universe is nearly flat at the beginning, there is 
Inflation only if $\theta_{0} \gg \mp$, that is, because of 
(\ref{equipartizione}), if its shear is very large. 
In the second case, where the energy tied to curvature is comparable
with scalar field potential energy, 
Universe inflate for any value of $\theta_{0}$ and $\sigma_{0}$.

\noindent  
We must note that $\theta_{\star}$ falls suddenly
at $\beta_{0} \approx \bs$. In fact, 
for $\beta_{0} \leq \bs$, the value of 
$\theta_{\star}$ is small ($\approx 10^{-3}$) but finite, whereas for 
$\beta_{0} > \bs$ $\theta_{\star}$ is null.
This is a consequence of strong non-linearity of Einstein's equations.
Slightly different initial conditions can evolve in completely different 
way.

\vspace{1cm}
\centerline{\bf 4. Conclusions}

\noindent
A strongly anisotropic Universe ($\theta \sim \sigma$) at early times, 
can go through an
Inflationary period, because of a scalar field minimally coupled to gravity.
In this case, equations(\ref{decadimento1})(\ref{decadimento2}) 
show that shear and curvature
became negligible with respect the expansion. In fact the Inflation 
causes the growth of $10^{40}$ times, at least, of linear dimension of the
Universe; this assure that $\sigma^{2}$ and $\curv$ decrease, respectively, of
$10^{240}$ and $10^{80}$ times. The expansion $\theta$, instead, 
remains nearly constant. In this case, at the end of Inflation, the Universe
can be described with the flat F.R.W. metric.

\noindent
For $\beta_{0} \geq \bs$ this situation happens for 
\underline{any} initial value of 
$\sigma$, $\theta$ and $\curv$; hence, even if the anisotropy and curvature
`energy' are very larger than scalar field `energy'. 
For $\beta_{0} \leq \bs$, instead, Universe does not inflate only if 
$\theta_{0} \leq \frac{\sqrt{3}}{3}\theta_{\star}$.

\noindent
The fundamental feature to underline is that the set of
initial condition that do not allow inflation is \underline{small} but 
\underline{finite}

\noindent
In the spirit of chaotic Inflation, we could
assume that any form of energy at Planck's epoch be of Planck's energy order,
$\sim \mp$ (i.e. the energy `equipartition' that Barrow has proposed). 
This would imply that Inflation would be avoided for 
$\beta_{0} \leq .883\bl$,
because $\theta_{\star} \sim \mp$ at $\beta_{0} \sim .883\bl$ and,
when $\beta_{0} \rightarrow \bl$, it reaches to $54.60 \mp$.

\noindent
In this simple model, then, neither the initial values of dynamical quantities
$\theta, \sigma$ nor the geometrical quantity $\curv$, 
would determine the evolution; the only important quantity would be $\beta$.

\vspace{.5cm}
\noindent
The authors are very grateful to dott.Francesco Chierici and 
dott.Carlo Bertoni for useful comments and suggestions.

\newpage

\begin{table}
\begin{center}
\begin{tabular}{|l|c|c|c|c|c|}
\hline
Model              &       &  H.C.    &B.J.S.         & B.F.S        
& M.E.S.       \\
\hline 
                   &       &          &               &              
&              \\  
Bianchi I          &$\di$  &$6\p10^{-8}$&               &              
&              \\
                   &       &          &               &
&              \\  \cline{1-5} 
                   &$\di$  &$1.5\p10^{-4}$ &               &              
&              \\
Bianchi V          &       &          &               &
&              \\
                   &$\ro$  &$4\p10^{-4}$ &$1.7\p10^{-7}$ &              
&              \\  \cline{1-5}
                   &$\di$  &$\xu 10^{-6}$ &               &              
&              \\
Bianchi $VII_{0}$  &       &          &               &
&              \\
                   &$\ro$  &$\xd 10^{-6}$ &$3.2\p10^{-9}$ &              
& $10^{-5}$    \\  \cline{1-5}  
                   &$\di$  &$\xd 10^{-3}$ &               &$0.5\p10^{-9}$
&              \\
Bianchi $VII_{h}$  &       &          &               &
&              \\
                   &$\ro$  &$\xu 10^{-3}$ &$1.2\p10^{-7}$ &$10^{-8}$     
&              \\  \cline{1-5}  
                   &$\di$  &$10^{-3}$ &               &              
&              \\
Bianchi IX         &       &          &               &
&              \\
                   &$\ro$  &$10^{-11}$&$3.9\p10^{-13}$&              
&              \\
\hline
\end{tabular}
\end{center}
\caption{\em LIMITS ON SHEAR ($\sigma$) AND VORTICITY ($\omega$)
OF THE UNIVERSE FROM C.M.B.R OBSERVATIONS.
In Bianchi $VII_{0}$ and Bianchi $VII_{h}$ models there is an 
adjustable parameter $x$; the ratio of the comoving length scale over which the
orientation of the principal axes of shear change, to the present Hubble
radius. 
In Bianchi $VII_{h}$ and Bianchi V open models,
we chose the limits for models with $\Omega_{0}=0.1$.
Hakwing,Collins (H.C.) calculated the upper limits on anisotropy in 1973,
when the upper limits on C.M.B.R. temperature anisotropy were 
approximately three order of magnitude greater than values observed by 
C.O.B.E.; their results, after correction, become comparable with the 
other ones. 
Barrow,Juszkiewicz,Sonoda (B.J.S) studied only the rotation, 
while Bunn,Ferreira,Silk (B.F.S) limited to Bianchi $VII_{h}$. Both studies
do not give analytic expression of the limits with respect to parameter
$x$, for the comparison we chose $x=1$
(the limits are those from the quadrupole anisotropy). 
The Maartens,Ellis,Stoeger (M.E.S) analysis, instead, does not depend on 
the model, and they have the same order of magnitude for shear and vorticity.}
\label{limiti}
\end{table}

\end{document}